\documentclass[twocolumn,tighten]{aastex63}
\usepackage{mathtools}
\usepackage{txfonts} 
\usepackage{hyperref}
\usepackage{color}

\usepackage{amsmath,amssymb,mathrsfs}

\def\be{\begin{eqnarray}}
\def\ee{\end{eqnarray}}
\def\ba{\begin{array}}
\def\ea{\end{array}}

\let\a=\alpha \let\b=\beta \let\g=\gamma \let\d=\delta \let\e=\epsilon
\let\h=\eta  \let\k=\kappa \let\l=\lambda 
\let\t=\tau \let\u=\upsilon
\let\r=\rho
 
\let\w=\omega	   \let\G=\Gamma \let\D=\Delta  \let\L=\Lambda
\let\S=\Sigma \let\W=\Omega
\let \x = \chi \let \z = \xi

\newcommand{\m}{M87*}
\newcommand{\s}{Sgr\,A*}

\begin{document}                                              

\title{The Structure in the Structure Function of Black Hole Light Curves}

\shorttitle{The Structure in the Structure Function}

\author[0000-0002-7179-3816]{Daniel~C.~M.~Palumbo}
\affil{Center for Astrophysics $\vert$ Harvard \& Smithsonian, 60 Garden Street, Cambridge, MA 02138, USA}
\affiliation{Black Hole Initiative at Harvard University, 20 Garden Street, Cambridge, MA 02138, USA}

\author[0000-0001-9528-1826]{Alejandro C\'ardenas-Avenda\~no}
\affiliation{Department of Physics, Wake Forest University, Winston-Salem, North Carolina 27109, USA}

\author[0000-0001-8986-5403]{Lena Murchikova}
\affiliation{Center for Interdisciplinary Exploration and Research in Astronomy, Northwestern University, Evanston IL 60208, USA}
\affiliation{Department of Physics \& Astronomy, Northwestern University, Evanston IL 60208, USA}
\affiliation{School of Natural Sciences, Institute for Advanced Study, 1 Einstein Drive, Princeton, NJ 08540, USA}

\begin{abstract}

Of the two black holes, M87$^\ast$ and Sgr~A$^\ast$, for which event-horizon-scale imaging is currently achievable, only Sgr~A$^\ast$ at our Galactic Center exhibits sub-hour variability, making light curve observations of the inner accretion flow especially valuable for time-domain studies of the spacetime and the surrounding plasma. Observed light curves are often sampled with variable cadence, so a structure function is often used to quantify their variability properties. In this work, we assemble a heterogeneous Sgr~A$^\ast$ light curve data set from discrete observations acquired over $\sim12$ years {at 230 GHz}, with different instruments, durations, and cadences, and study its structure function in three frameworks. First, we consider analytic predictions for the features of structure functions of stationary random processes with Fourier spectra dictated by the Mat\'ern covariance. We then study the structure functions resulting from lensed movies of synthetic optically thin accretion flows approximated with Mat\'ern covariance in both space and time, and identify which features in the structure function can be attributed to the black-hole geometry. Lastly, we examine the structure functions of a library of $72$ general relativistic magnetohydrodynamic simulations. We find that, in general, magnetically arrested disk (MAD) models produce smoother structure functions with steeper slopes than standard and normal evolution (SANE) counterparts. On time scales of $5-15$ minutes, the observed structure function slope of Sgr~A$^\ast$ more closely aligns with the MAD trend, while both simulation types are generally more variable than the data on all sub-hour timescales.

\end{abstract}


\section{Introduction}

\begin{figure*}[ht!]
    \centering
    \includegraphics[width=0.99\linewidth]{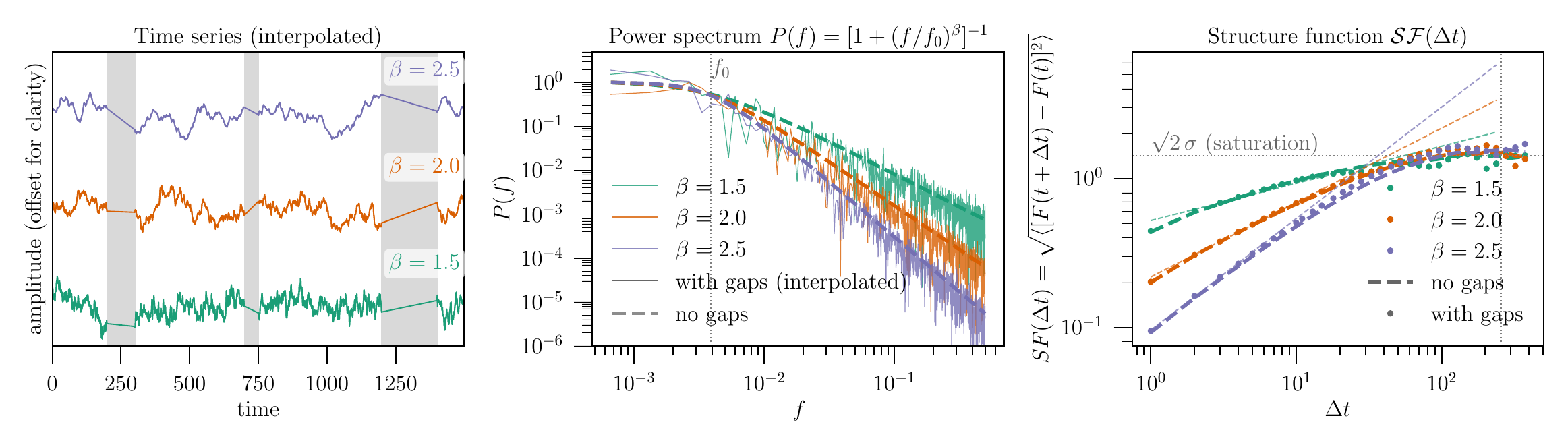}
    \caption{Example unitless power-law-spectrum processes {with power law index $\beta$} with gaps in sampling (left) as well as the corresponding measured power spectral density (middle) and structure function (right). After gaps are introduced, the time series is naively linearly interpolated. Small-lag structure is nearly unchanged by the data gaps, whereas the {gaps introduce broad-spectrum noise to the power spectral density.}}
    \label{fig:pedagogy}
\end{figure*}

For unresolved astrophysical sources, one of the primary observables at a given frequency is the so-called ``light curve,'' which is the time-dependent total flux density received from the source. In the case of interferometric observations, it can also be the total correlated flux density measured on a particular spatial scale in the source. With the exception of a few special cases, the central engines of black holes together with their accretion flow are unresolved sources. In the context of black holes and their accretion flows, 
light curves probe near-horizon dynamical timescales.

In this work, we focus on supermassive black holes (SMBHs), in particular, low-luminosity active galactic nuclei, and the prototypical and best-studied example of such an SMBH is the Milky Way's central black hole, Sagittarius A* (\s{}). For such black holes, the accretion disks are expected to be hot, geometrically thick, and dominated by advection, so the innermost stable circular orbit is expected to play a less direct role than in thin-disk systems \citep{Ichimaru1977, Rees_1982, Narayan_1994, Reynolds_1996}. {As a result, rather than coherent orbital motions, red noise-like variability is expected from realistic hot accretion disks \citep[see, e.g.,][]{SgrA_PaperV}.} 

Recent Event Horizon Telescope (EHT) studies of Sgr~A$^\ast$ quantified the variability of the horizon-scale emission and compared observed light curves with libraries of general relativistic magnetohydryodynamic (GRMHD) models. Such simulations are commonly classified by their magnetic state. In magnetically arrested disk (MAD) models, accumulated magnetic flux becomes dynamically important and can intermittently impede or regulate accretion \citep{Ichimaru1977,Igumenschchev_2003,Narayan2003}. {Meanwhile}, in standard and normal evolution (SANE) models, the net magnetic flux is comparatively weaker, and the flow is less magnetically arrested \citep{Narayan_2012}. These two classes exhibit different light-curve variability, yet many torus-fed models do not satisfy all observational constraints, motivating variability diagnostics beyond a single fluctuation amplitude \citep{SgrA_PaperIV,SgrA_PaperV}.

While simulated light curves are typically evenly sampled and can be analyzed using standard spectral methods, SMBH observational data tend to be unevenly sampled. SMBH variability is relatively slow, and observations are usually composed of independent datasets from different telescopes with different cadences. For unevenly sampled light curves, structure functions provide a natural way to quantify variability \citep[e.g.,][]{Simonetti1985,Do2009ApJ,Yusef-Zadeh2011,Dexter2014,Witzel2021,Murchikova_2022,Wielgus2022}\footnote{Particularly extensively structure functions are used in physics of turbulence, which dates back to \cite{Kolmogorov1941}.}. As an example, \autoref{fig:pedagogy} shows how a simple deletion of data (followed by linear interpolation over the gaps) leaves the short lag behavior of the structure function essentially unchanged, while the power spectral density (PSD) suffers a broadband corruption from the hard edges associated with each gap. In both cases, the long-lag and low-frequency structure also scatter from the smooth white-noise prediction as a result of the data deletion and interpolation.

In this work, we study the SMBH variability properties using the square root of the second-order structure function, which measures the RMS flux difference between pairs of points on light curves separated by a time lag $\Delta t$. The standard expression for the structure function ($SF$) is 
\begin{equation}
\label{eq:SF_estimator_noiseless}
SF(\Delta t)
=
\sqrt{
\frac{1}{N_{\Delta t}}
\sum_{\textrm{pairs}}
\big[F(t+\Delta t)-F(t)\big]^2
}\,,
\end{equation}
where $F(t)$ is the source flux at time $t$, $N_{\Delta t}$ counts all pairs separated by $\Delta t$ within the time series, and the summation runs over all pairs of $F(t)$ and $F(t+\Delta t).$ This expression is suitable for calculating the structure function of noiseless data, like the data from numerical simulations. 

However, in the case of observations, an additional observational noise, uncorrelated with the source flux, tends to be present. Atmospheric fluctuations, thermal noise within telescope components, and related instrumental effects are all considered observational noise here. Luckily, such uncorrelated noise can be easily removed {when present in the structure function as a white-spectrum short-lag plateau} \citep{Murchikova2021}. The following expression for the noise-corrected structure function $(\mathcal{SF})$ can be used instead\footnote{Some authors of observational variability studies tend to present structure functions without subtracting the observational-noise contribution (i.e., $SF(\Delta t)$ not $\mathcal{SF}(\Delta t)$). In such cases, the structure function plateaus at short $\Delta t,$ where observational noise dominates the signal. At lags where the plateau occurs, the $SF(\Delta t)$ does not reflect the properties of the source but only the properties of the uncorrelated observational noise.}:
\begin{equation}
\mathcal{SF}(\Delta t)
=
\sqrt{
\frac{1}{N_{\Delta t}}
\sum_{\textrm{pairs}}
\big[F(t+\Delta t)-F(t)\big]^2
-
2\sigma_{\rm obs}^2
}\,,
\label{eq:SF_estimator}
\end{equation}
where $\sigma_{\rm obs}$ is the RMS observational uncertainty {dominant at short lags; indeed, the $2\sigma_{\rm obs}^2$ term is simply the smallest-lag value of $SF(\Delta t)^2$}. In the text below, we use the term structure function to describe $\mathcal{SF}(\Delta t)$. Note that for processes with no observational/thermal noise component, such as those generated from toy models or numerical simulations, we have $\mathcal{SF}(\Delta t)= SF(\Delta t).$ {Provided that the underlying random process is stationary (though not necessarily zero-mean), $\mathcal{SF}^2 \approx 2  \sigma_{\rm \Delta t}^2 [1-C(\Delta t)]$, where $\sigma_{\Delta t}^2$ is the variance of the process on samples separated by a time $\Delta t$, and $C$ is the autocorrelation at that lag. Black hole accretion flows may exhibit secular trends in accretion rate that violate the assumption of stationarity. More acutely, sources with a particularly red variability spectrum may show weak trends in the mean over individual observations.} Note that some authors prefer to use the structure function in squared units. We instead use $\mathcal{SF}^2(\Delta t)$ to refer to the squared quantity. 

Throughout this work, we will touch on the time-delayed echoes of light associated with multiply lensed copies of emission near a black hole reaching the same observer after differing numbers of half-orbits around the black hole, counted by an index $n$ \citep{Johnson_2020}. This phenomenon is well studied \citep[see, e.g.,][]{Bardeen_1973, Cunningham_Bardeen_1973} and has been targeted for observations in light curve autocorrelations \citep{Chesler_2021, Hadar_2021, Wong_2024, CardenasAvendano2024}. We will discuss the time-domain manifestation of the same feature, which would (for a single emitter) generally manifest as a reduction in the structure function value at the light echo time. 

This Letter is organized as follows. 
In \autoref{sec:obs_light_curve}, we review the observed 230 GHz light curves of \s{}, which are the focus of this work and which we compare to a number of models throughout the Letter. In \autoref{sec:relativistic_effects}, we discuss how relativistic effects from the black hole itself contribute to the light curve observed from a turbulent plasma. In \autoref{Sec:GRMHDSims}, we examine the structure function observed in black hole accretion disk simulations and find notable differences in the short-lag slopes between disks in different magnetization states. We conclude in \autoref{sec:discussion}.

\section{Observational Light Curves}
\label{sec:obs_light_curve}

We constructed the observational structure function $\mathcal{SF}(\Delta t)$ of \s{} at 230 GHz, near the peak of its spectral energy distribution, using published observations from the Atacama Large Millimeter/submillimeter Array (ALMA), the Submillimeter Array (SMA), and the Combined Array for Research in Millimeter-wave Astronomy (CARMA) \citep{Dexter2014,Witzel2021,Murchikova2021,Wielgus2022,Wenner_2026}. The data span 2009--2021.
The highest cadence in the simulated light curves is $5M$ (Section~\ref{Sec:GRMHDSims}), corresponding to $\approx100$ s for \s, assuming the mass of $4.1\times10^6\,M_\odot$. We therefore evaluate the observational structure function $\mathcal{SF}(\Delta t)$ only for $\Delta t \ge 100$ s. 

We combine all observational data into a single time series, compute $(F(t+\Delta t)-F(t))^2$ for all possible pairs $F(t_1)$ and $F(t_2)$, bin the results by $\Delta t$, evaluate Eq.~\ref{eq:SF_estimator} in each bin, and assign the obtained value to the center of the $\Delta t$-bin. We approximate the observational-noise contribution as $ 2\sigma_{\rm obs}^2 \simeq \mathcal{SF}(\Delta t_{\min}\simeq 4 \textrm{s}), $ where $\Delta t_{\min} \ll 100$ s, and thus, this approximation is fairly accurate. 

A more complex procedure was used by, for example, \citet{Murchikova_2022}, who added light curves gradually based on their individual observational uncertainties. This improves the accuracy of $\mathcal{SF}(\Delta t)$ on timescales of a few seconds, but has little effect for $\Delta t > 100\,\mathrm{s}$. We therefore use the simpler approach, which is easier to reproduce.

\begin{figure}[t!]
    \centering   \includegraphics[width=0.99\columnwidth]{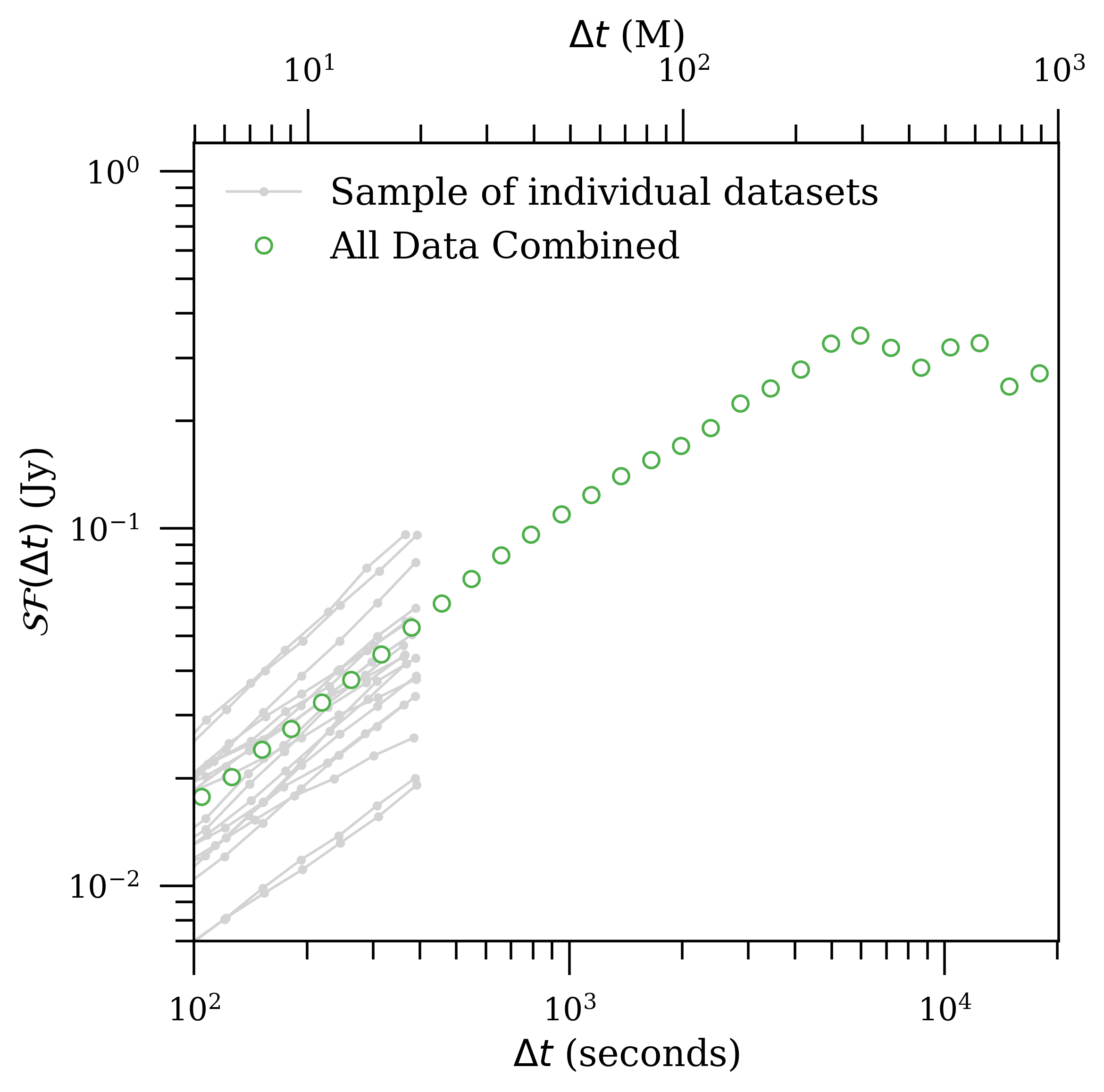}
    \caption{Observational structure function of \s{} obtained from the combined ALMA, SMA, and CARMA data described in Section~\ref{sec:obs_light_curve} (green circles). Gray lines show structure functions from several individual ALMA light curves, included to illustrate the scatter among observing epochs. ALMA data are used because they have the highest signal-to-noise ratio. The individual structure functions have mostly similar slopes but different normalizations. They are shown only for $\Delta t \leq 400\,\mathrm{s}$, approximately one-tenth of the individual light-curve durations, where the statistics remain robust. Time lags are shown in both seconds and gravitational time units for the Galactic Center supermassive black hole.}
    \label{fig:ObsData}
\end{figure}

As shown in Figure \ref{fig:ObsData}, the combined time series allows the evaluation of the structure function on lags as high as $10^4$ seconds, at which the apparent break in the power law structure can indicate the transition to white-noise-like decorrelated behavior. However, at these large lags, there is a low effective number of realizations, and thus, we do not think this break is a reliable indicator of the transition to white-noise asymptotic behavior (shown in the top right of Figure \ref{fig:pedagogy}).

\section{Relativistic Contributions to the Structure Function}
\label{sec:relativistic_effects}

In astrophysical accretion flows, the emitting region is expected to naturally provide a continuum of characteristic timescales that vary with position. Consequently, the structure function of a black hole accretion flow light curve is not expected to exhibit intrinsically sharp breaks but should instead exhibit smooth bends, extended scaling regions, or broad knees. The effective range of lags and the prominence of different timescales depend on which parts of the flow contribute most to the observed flux. This dependence is \emph{jointly} set by plasma physics and the general relativistic effects on light propagation. For the former, the plasma supplies intrinsic variability through the statistical properties of the emission region (e.g., correlation times, spatial coherence, and radial stratification). For the latter, the propagation of the emitted light through the black hole spacetime reweights that variability through Doppler shifts, relativistic aberrations, and light bending.

In this section, we study how relativistic effects shape the structure function of black hole emission. Note that here, $\mathcal{SF} (\Delta t) = SF(\Delta t)$ as there is no observational noise $\sigma_\mathrm{obs}=0$ in this system. So we can omit the noise term in Eq.~\eqref{eq:SF_estimator}, and use Eq.~\eqref{eq:SF_estimator_noiseless}. We will separate the relativistic contributions by first discussing the \emph{direct} image contribution ($n=0$), and then adding the delayed, demagnified contribution from the photon ring, using a simplified model.  

\subsection{Direct image contribution ($n=0$)}
\label{sec:direct_image}

Let $\mathcal{G}(r_{\rm s},\phi_{\rm s},t_{\rm s})$ be the source-plane emission field. We consider emission confined to the equatorial plane, with $(r_{\rm s},\phi_{\rm s},t)$ denoting the coordinates on the emission plane. The intrinsic (unlensed) integrated light curve is
\begin{equation}
\mathscr{L}_{\rm s}(t)
\equiv
\int_{r_{\rm in}}^{r_{\rm out}}\!\!\int_{0}^{2\pi}
\mathcal{G}(r_{\rm s},\phi_{\rm s},t)\,dr_{\rm s}\,d\phi_{\rm s},
\label{eq:Ls_def}
\end{equation}
with autocorrelation $C_{\rm s}(\Delta t)\equiv
\langle\mathscr{L}_{\rm s}(t)\mathscr{L}_{\rm s}(t+\Delta t)\rangle$. For a stationary process with lag-only autocorrelation, the corresponding (noise-free) structure
function satisfies $\mathcal{SF}_{\rm s}^2(\Delta t)=2\,[C_{\rm s}(0)-C_{\rm s}(\Delta t)]$. This light curve $\mathscr{L}_{\rm s}(t)$ is not directly observable, but it is useful for understanding the relativistic effects caused by the propagation of photons. 

The direct lensed contribution ($n=0$), i.e., the emission that reaches the observer but only from photons that do not execute a full orbit around the black hole, can be written as a linear functional of the intrinsic emission field,
\begin{equation}
\mathscr{L}_{0}(t)
=
\int_{r_{\rm in}}^{r_{\rm out}}\!\!\int_{0}^{2\pi}
W(r_{\rm s},\phi_{\rm s})\,\mathcal{G}(r_{\rm s},\phi_{\rm s},t)\,dr_{\rm s}\,d\phi_{\rm s},
\label{eq:L0_general_weight}
\end{equation}
where the weight $W(r_{\rm s},\phi_{\rm s})$ encodes the mapping from the source plane to the signal received by the observer. In full ray tracing, $W$ includes (i) the redshift factor $g^3$ from specific-intensity invariance, $I_{\rm o}=g^3 I_{\rm s}$, and (ii) the Jacobian relating source-plane area elements to the solid angle on the observer's screen, which plays a role analogous to the transfer function in disk-line calculations
(e.g.,~\citealt{Cunningham1975}). Eq~\eqref{eq:L0_general_weight} illustrates the key point: relativistic propagation does not generate new intrinsic variability, but it changes \emph{which parts} of the flow dominate the observed light curve. 

To connect Eq.~\eqref{eq:L0_general_weight} to the morphology of the structure function, let us further reduce the emission field (disk) to a set of (weakly correlated) annuli. We define the azimuth-integrated annular signal
\begin{equation}
J(r_{\rm s},t)\equiv \int_{0}^{2\pi}\mathcal{G}(r_{\rm s},\phi,t)\,d\phi,
\label{eq:J_def}
\end{equation}
and assume that different radii contribute approximately independently (after azimuthal integration), i.e., 
\begin{equation}
\left\langle J(r_{\rm s},t)\,J(r'_{\rm s},t+\Delta t)\right\rangle
\simeq
\delta(r_{\rm s}-r'_{\rm s})\,\mathcal{C}_J(r_{\rm s};\Delta t),
\label{eq:annulus_closure_CJ}
\end{equation}
where $\mathcal{C}_J(r;\Delta t)\equiv \langle J(r_{\rm s},t)J(r_{\rm s},t+\Delta t)\rangle$.
Under these assumptions,
\begin{equation}
\mathscr{L}_{\rm s}(t)=\int_{r_{\rm in}}^{r_{\rm out}}J(r_{\rm s},t)\,dr_{\rm s},
\quad
\mathscr{L}_{0}(t)=\int_{r_{\rm in}}^{r_{\rm out}}w(r_{\rm s})\,J(r_{\rm s},t)\,dr_{\rm s},
\label{eq:Ls_L0_annuli_w}
\end{equation}
where $w(r_{\rm s})$ is the azimuth-averaged effective weight induced by propagation. The corresponding autocorrelations are one-dimensional radial integrals,
\begin{align}
C_{\rm s}(\Delta t)&=\int_{r_{\rm in}}^{r_{\rm out}}\mathcal{C}_J(r_{\rm s};\Delta t)\,dr,
\\ 
C_{0}(\Delta t)&=\int_{r_{\rm in}}^{r_{\rm out}}w^2(r_{\rm s})\,\mathcal{C}_J(r_{\rm s};\Delta t)\,dr_{\rm s},
\label{eq:Cs_C0_annuli_w}
\end{align}
and the direct-image structure function follows as
\begin{equation}
{\mathcal{SF}_{0}}^2(\Delta t)
=
2\int_{r_{\rm in}}^{r_{\rm out}}w^2(r_{\rm s})\Big[\mathcal{C}_J(r_{\rm s};0)-\mathcal{C}_J(r_{\rm s};\Delta t)\Big]\,dr_{\rm s}.
\label{eq:SF0_annuli_w}
\end{equation}
Equations~\eqref{eq:Cs_C0_annuli_w} and~\eqref{eq:SF0_annuli_w} explicitly show that direct-image relativistic effects reshape the observed structure function $\mathcal{SF}_0$ through a
deterministic reweighting of intrinsic annular correlations. They thereby change the effective mixture of short and long timescales contributing to the observed signal. 

This expression implies the following qualitative behavior. At small lags, $\mathcal{SF}_0(\Delta t)$ is controlled by the shortest decorrelation times present in the \emph{weighted} mixture selected by $w^2$. For lags large enough that the light curve has decorrelated, $\mathcal{SF}_0^2(\Delta t)$ saturates at $2C_0(0)=2\,\mathrm{Var}[\mathscr{L}_0]$. The plateau height therefore depends on how strongly the illumination pattern concentrates the contribution to the observed flux (a more concentrated illumination pattern produces less self-averaging, increasing the variance and hence the plateau height). The knee and any extended scaling region are set by the distribution of intrinsic correlation times across the disk after reweighting.

\subsection{Controlled demonstrations with the \texttt{inoisy} and \texttt{AART} codes}
\label{sec:inoisy_aart_demo}

We now show how the aforementioned model provides the main ingredients that shape the structure function using the \texttt{AART} \citep{CardenasAvendano2023} and \texttt{inoisy} \citep{Lee2021} codes. The \texttt{inoisy} code provides controlled source-plane emissivity fluctuations by generating a Gaussian random field (GRF) $\mathcal{F}(\mathbf{x})$ as the solution of the stochastic partial differential equation (SPDE) \citep{Lee2021}:
\begin{equation}
\left(1-\nabla\cdot\mathbf{\Lambda}(\mathbf{x})\nabla\right)\mathcal{F}(\mathbf{x})
=
\mathcal{N}\,|\mathbf{\Lambda}(\mathbf{x})|^{1/4}\,\mathcal{W}(\mathbf{x}),
\label{eq:inoisy}
\end{equation}
where $\mathcal{W}(\mathbf{x})$ is a white-noise process, $\mathcal{N}$ is a normalization constant, and the correlation structure is encoded in
\begin{equation}
\mathbf{\Lambda}(\mathbf{x})=\sum_{\ell=1}^{d}\lambda_\ell^2(\mathbf{x})\,
\mathbf{u}_\ell(\mathbf{x})\,\mathbf{u}_\ell^{\rm T}(\mathbf{x}),
\qquad
|\mathbf{\Lambda}|\equiv\det \mathbf{\Lambda}=\prod_{\ell=1}^{d}\lambda_\ell^2.
\label{eq:CorrelationStructure}
\end{equation}
Here $\{\mathbf{u}_\ell\}$ are orthonormal directions, and $\{\lambda_\ell\}$ are the correlation lengths. Spatial inhomogeneity and anisotropy follow when the $\lambda_\ell$ are unequal or the $\mathbf{u}_\ell$ vary across the domain. In the present context, changing the $\lambda_\ell$ (or their functional dependence
on radius) amounts to changing the \emph{astrophysical} content of the model, i.e.,  the ``astrophysics'' enters statistically through the assumed correlation structure and the spatial distribution of variability power rather than through a first-principles plasma calculation. 

The GRF is now mapped to a strictly positive source intensity by applying an envelope $f(r_{\rm s})$ and a lognormal transformation~\citep{CardenasAvendano2023}
\begin{equation}
I_{\rm s}(r_{\rm s},\phi_{\rm s},t)
=
f(r_{\rm s})\,
\exp\!\left[\sigma_{\rm scale}\,\mathcal{F}(r_{\rm s},\phi_{\rm s},t)-\frac{1}{2}\sigma_{\rm scale}^2\right],
\label{eq:envelope}
\end{equation}
where $\sigma_{\rm scale}$ controls the fluctuation amplitude. The envelope is a proxy for the mean emissivity profile: it reweights which radii contribute most strongly to the integrated light curve, and thus can modify the observed PSD of the resulting field and, therefore, also the structure function in a manner qualitatively similar to propagation weights. In general, both the astrophysical weighting (through $f$ and the intrinsic correlation structure) and the relativistic transfer (through $W$) act by selecting and re-weighting a hierarchy of intrinsic timescales. 

In the simplest homogeneous case (constant $\lambda_\ell$ and constant $\mathbf{u}_\ell$), the resulting GRF is Mat\'ern and its Fourier-space covariance is \citep{CardenasAvendano2023}
\begin{equation}
\label{eq:MaternCovariance}
	\widetilde C_\nu(k)= \frac{(4\pi)^{d/2}\Gamma(\nu+d/2)\lambda^d} {\Gamma(\nu)\,[1+\lambda^2k^2]^{\nu+d/2}},
\end{equation}
where $d$ is the dimension of the field and $\nu=2$ is the covariance parameter. Given a realization of $I_{\rm s}$, we compute the unlensed light curve $\mathscr{L}_{\rm s}(t)$ by integrating $I_{\rm s}$ over the source plane, and then ray trace, in the slow-light mode, the same emissivity through the spacetime with the \texttt{AART} code~\citep{CardenasAvendano2023} to obtain the lensed light curve and its decomposition into image orders, including the direct component $\mathscr{L}_0(t)$. 

\begin{figure*}
\centering
\includegraphics[width=0.99\textwidth]{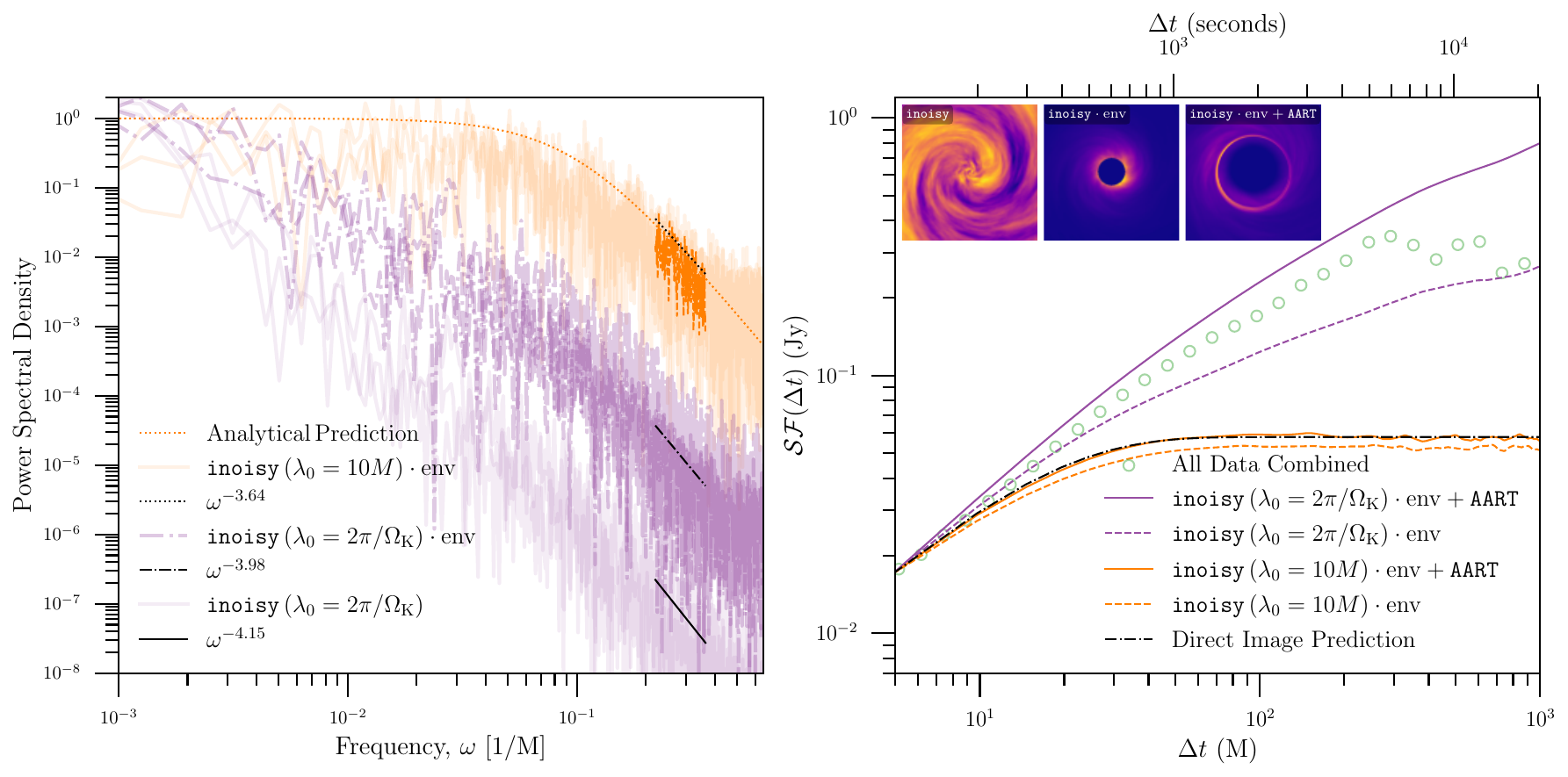}
\caption{Power spectral densities (PSDs) and structure functions for controlled \texttt{inoisy}+\texttt{AART} experiments. \emph{Left:} PSDs for two \texttt{inoisy} prescriptions. Orange curves correspond to a constant time correlation $\lambda_0=10\,M$, while purple curves correspond to a Keplerian-like prescription $\lambda_0(r)=2\pi/\Omega_{\rm K}(r)$. In each case, solid lines correspond to the ray-traced light curves (\texttt{inoisy}+\texttt{AART}), dashed lines correspond to the emissivity model including the envelope only
(\texttt{inoisy}), and dotted lines correspond to the underlying GRF (no envelope). The short black line segments indicate power-law fits to the PSD slope over the frequency ranges shown. \emph{Right:} the corresponding structure functions of the cases shown in the left panel. Green open circles show the structure function computed from Sgr~A$^\ast$ observations (see Sec.~\ref{sec:obs_light_curve}). The simulated structure functions are normalized to the observational one to emphasize differences in morphology. The inset shows snapshots of the GRF, the enveloped \texttt{inoisy} emissivity, and the resulting ray-traced image. The black dash-dotted curve shows the analytic lensing contribution for the constant $\lambda_0=10\,M$ case.}
\label{fig:psd_sf_inoisyAART}
\end{figure*}

We can now use this semi-analytical model to exemplify and understand the effects of lensing described in the previous section. The left panel of Figure~\ref{fig:psd_sf_inoisyAART} shows the resulting power spectrum of light curves produced from this model. As noted in \cite{Lee2021}, for the homogeneous case ($\lambda_\ell=\rm{const.}$), due to the envelope and the noise realization, the resulting PSD does not necessarily follow the simple behavior shown in Eq.~\ref{eq:MaternCovariance}, i.e., a decay proportional to $\omega^{-4}$ (plotted with a dotted line).  In the left panel of Figure~\ref{fig:psd_sf_inoisyAART} for each case of the time correlation $\lambda_0$, three different noise realizations are presented, and we have performed a fit to the post-knee decay of the realization-averaged PSD. As expected for this case, the envelope does not substantially change the temporal PSD slope in this homogeneous case, although it changes the normalization and spatial weighting. However, in the inhomogeneous and anisotropic case ($\lambda_\ell \neq \rm{const.}$, as depicted by the purple lines in the figure), the envelope plays a crucial role. It can modify the power spectrum's steepness or shallowness, contingent upon the functional form of the envelope. This demonstrates the model's versatility in generating diverse power spectra. 

The right panel of Figure~\ref{fig:psd_sf_inoisyAART} summarizes the effect of both the intrinsic prescription and relativistic propagation. It shows the structure function computed from Sgr~A$^\ast$ observations, the green open circles described in Sec.~\ref{sec:obs_light_curve}, together with synthetic structure functions normalized to the observations computed with \texttt{AART}. The orange dashed curve is the structure function computed directly from Eq.~\ref{eq:envelope} with a constant correlation time $\lambda_0=10\,M$; the orange solid curve shows the same emission after ray tracing with \texttt{AART}. Ray tracing changes the morphology, as expected from the relativistic reweighting discussed above.

Given that the constant-correlation-time case ($\lambda_0=10\,M$) is analytically tractable, we can develop further intuition and understand this change. In the $g^3$-dominated limit, this implies $w^2(r)\simeq g^6(r+1)$. If we consider the Schwarzschild spacetime ($a=0$), a face-on observer ($i=0^\circ$), and material moving on circular geodesic motion, the redshift factor is $g=\sqrt{1-3/r}$, and one can use the proxy $r_{\rm o}\approx r_{\rm s}+1$ for the direct-image mapping (e.g.,~\citealt{GrallaLupsasca2020}). Even though these are quite rough assumptions, they remarkably predict the resulting (relativistic) structure function, as demonstrated by the black dash-dotted curve in the right panel of Figure~\ref{fig:psd_sf_inoisyAART}. This prediction was made using Eq.~\ref{eq:SF0_annuli_w} and the aforementioned assumptions. 

On the other hand, when the flow provides a continuous range of correlation times (as anticipated when characteristic times vary with radius), $w^2$ becomes more intricate. This can cause the effective knee to shift by either enhancing or suppressing radii associated with faster or slower intrinsic variability, in a non-analytic but numerically tractable manner, {acting as a smooth frequency-domain filter on individual sub-images governed by relative contributions from individual radii}. An example of this behaviour can be seen from the purple lines in the right panel of Figure~\ref{fig:psd_sf_inoisyAART}, where the dashed curve corresponds to a source-plane \texttt{inoisy} realization with a radially varying correlation time, $\lambda_0(r)=2\pi/\Omega_{\rm K}(r)$, representing a Keplerian-like ($\Omega_{\rm K}(r) \propto r^{-3/2}$) hierarchy of intrinsic timescales; this curve initially coincides with the constant-$\lambda_0$ model at short lags and then continues to rise as larger radii contribute longer memory. The purple solid curve shows the Keplerian \texttt{inoisy} realization after ray tracing, which follows the observational morphology more closely. We emphasize that these curves are not fits to the data; rather, they illustrate that (i) a Keplerian-like stratification of intrinsic correlation times is a reasonable first approximation and (ii) relativistic transfer seems to be \emph{required} to reproduce the observed structure function morphology. 

\begin{figure}
\centering
\includegraphics[width=0.99\columnwidth]{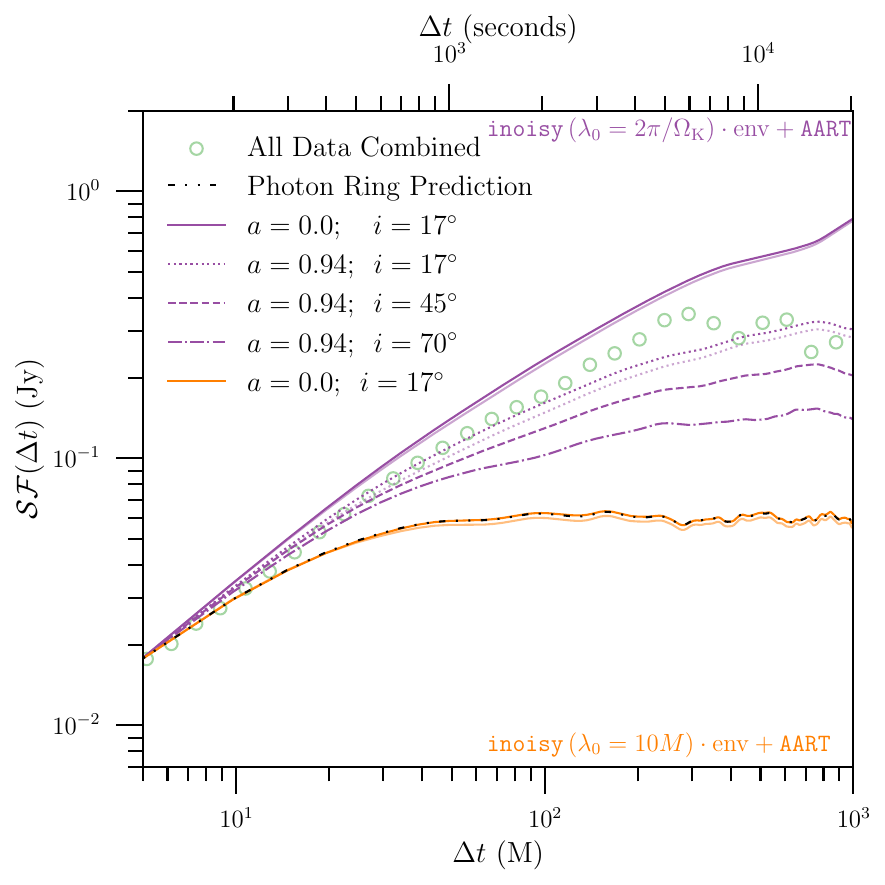}
\caption{Dependence of the structure function on spin and inclination for a fixed astrophysical prescription. The purple curves correspond to the Keplerian-like \texttt{inoisy} model with $\lambda_0(r)=2\pi/\Omega_{\rm K}(r)$ ray traced with \texttt{AART}, while the orange curve corresponds to the homogeneous case $\lambda_0(r)=10$. The lighter-colored curves correspond to the ray tracing of only the direct image ($n=0$). For the homogeneous case, we show the analytic photon ring contribution using $\tau\sim16.3\,M$ and $\gamma\sim0.9\pi$, values obtained from theory for $a=0.0$ and $i=17^\circ$. As we can see, varying the inclination and spin changes the morphology of the structure function. Curves are normalized as in Figure~\ref{fig:psd_sf_inoisyAART}, i.e., to Sgr~A$^\ast$ observed data, to highlight shape differences.}
\label{fig:spin_incl_sf}
\end{figure}

Because the dependence on spacetime geometry, black hole spin, and observer inclination is incorporated through the relativistic transfer function, the effective weight will not be purely radial and analytically tractable for generic plasma motion. Figure~\ref{fig:spin_incl_sf} shows how the structure function changes when varying $a$ and $i$ for a fixed astrophysical Keplerian-like prescription. As expected from the lensing behavior, these examples show that the inclination produces a larger change in the structure function morphology than spin. This supports the interpretation that, while relativistic effects are required to reproduce the detailed morphology from a Keplerian-like emission prescription, changes to the astrophysical variability model (e.g., through the distribution of intrinsic timescales and emissivity weighting) are typically the dominant source of variation, with inclination acting as the most important relativistic control
parameter among those explored here.

\subsection{The Photon Ring contribution to the Structure Function}
\label{sec:photon_ring}

Beyond the direct image, general relativity predicts that higher-order photon ring images can be approximated as delayed and demagnified copies of the direct light curve. In the previous section, we computed the resulting structure function accounting for all these contributions. In this section, we will focus on the first photon ring contribution, $n=1$, to the light curve. Following \citet{CardenasAvendano2024}, let us model the observed light curve as a sum over image order,
\begin{equation}
\mathscr{L}(t_{\rm o})
=
\sum_{n=0}^{\infty}\mathscr{L}_n(t_{\rm o})
\approx
\sum_{n=0}^{\infty}e^{-n\gamma}\,\mathscr{L}_0(t_{\rm o}-n\tau),
\label{eq:LightCurve}
\end{equation}
where $\tau$ is the characteristic delay between successive photon-ring orders and $\gamma$ controls the
exponential demagnification~\citep{GrallaLupsasca2020}. Retaining only the first correction ($n\le 1$) gives
\begin{equation}
\mathscr{L}(t)\simeq \mathscr{L}_0(t)+e^{-\gamma}\,\mathscr{L}_0(t-\tau).
\label{eq:L_trunc}
\end{equation}
Denoting the direct-image autocorrelation by
\begin{equation}
C_0(\Delta t)\equiv \left\langle \mathscr{L}_0(t)\,\mathscr{L}_0(t+\Delta t)\right\rangle,
(Cs(\Delta t)\equiv\langle\delta\mathcal Ls(t)\,\delta\mathcal Ls(t+\Delta t)\rangle
\label{eq:C0_def}
\end{equation}
the corresponding structure function can be written, to leading order in $e^{-\gamma}$, as
\begin{equation}
\mathcal{SF}^2(\Delta t)
\simeq
{\mathcal{SF}_0}^2(\Delta t)
+
2e^{-\gamma}\Big[2C_0(\tau)-C_0(\Delta t+\tau)-C_0(\Delta t-\tau)\Big].
\label{eq:SF_result}
\end{equation}
This expression makes explicit the contributions to the observed structure function: the intrinsic-plus-direct term $\mathcal{SF}_0(\Delta t)$, and another relativistic correction suppressed by $e^{-\gamma}$ that introduces a characteristic feature at lags of order $\tau$ through shifted evaluations of the autocorrelation. For $\Delta t\ll\tau$, a Taylor expansion gives
$C_0(\Delta t\pm\tau)=C_0(\tau)\pm \Delta t\,C_0'(\tau)+\tfrac12\Delta t^2 C_0''(\tau)+\cdots$, so the linear terms cancel, and the correction is controlled by the local curvature of the autocorrelation at $\tau$, $2C_0(\tau)-C_0(\Delta t+\tau)-C_0(\Delta t-\tau)\simeq -\Delta t^2 C_0''(\tau)$. Thus, at small lags, the photon-ring term produces only a smooth, subleading modification whose amplitude is suppressed by $e^{-\gamma}$. Around $\Delta t\simeq\tau$, the shifted arguments probe $C_0$ near $0$ and $2\tau$, yielding $2C_0(\tau)-C_0(0)-C_0(2\tau)$; since typically, $C_0(0)$ is the maximum, this can produce the largest relative distortion, potentially imprinting a kink, dip, or mild bump near the delay scale. Finally, for $\Delta t\gg\tau$, astrophysical light curves are expected to decorrelate so that $C_0(\Delta t\pm\tau)\to 0$, and the photon-ring contribution saturates to an additive constant.

For the constant-$\lambda_0$ Mat\'ern case, $C_0(\Delta t)$ is analytic, allowing a direct evaluation of Eq.~\eqref{eq:SF_result}. The black dash-dotted curve in the right panel of Figure~\ref{fig:spin_incl_sf} shows this analytic photon-ring prediction for the constant-$\lambda_0=10\,M$ model using $\tau=16.3\,M$ and $\gamma=\pi\times 0.9$, values obtained from theory for $a=0$ and $i=17^\circ$. Therefore, the dominant relativistic imprint on the structure function generally arises from the direct image, while higher-order photon-ring contributions enter as a suppressed correction, which is likely to be difficult to disentangle from the rest of the contributions. 

\section{Sgr A* GRMHD Light Curves}
\label{Sec:GRMHDSims}

\begin{figure*}[t!]
    \centering
    \includegraphics[width=0.99\textwidth]{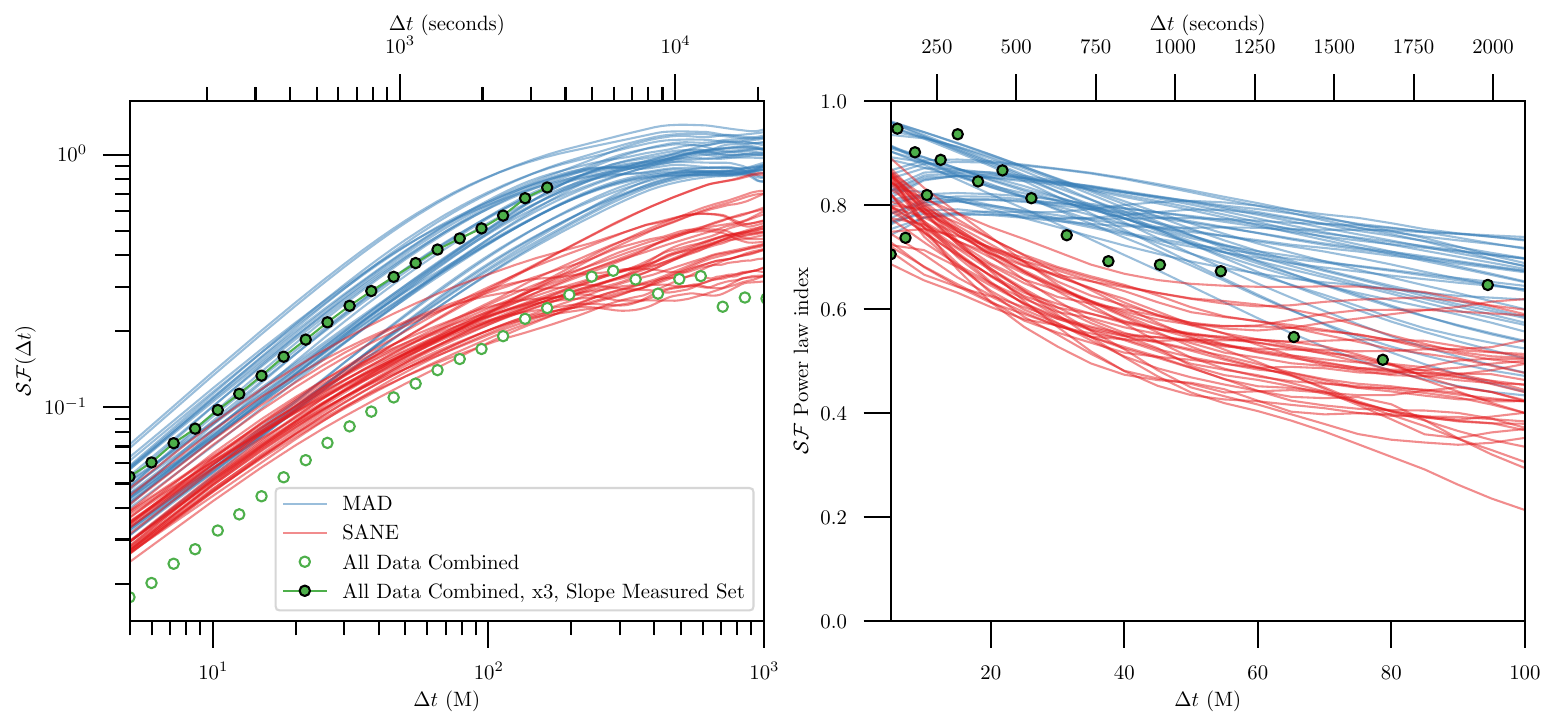}
    \caption{Comparison of the structure functions of MAD and SANE GRMHD simulations from the \s{} simulation library of $72$ movies. The open green circles show the measured observational structure function, while the filled green circles with black edges show the same values multiplied by three solely to compare shape; this constant factor does not affect the logarithmic slope. Left: structure functions evaluated at $300$ lags between $5$ and $1500$M. Right: slope of the log-log structure function (equivalently, the power law index of the structure function) evaluated on a sliding pair of points over the first $20$ lags. Evidently, on lags of hundreds of seconds for Sgr~A$^\ast$, SANE models bend to a shallower slope, suggesting the spectrum of Sgr~A$^\ast$ variability is a closer match to MAD dynamics.}
    \label{fig:GRMHD}
\end{figure*}

We now turn to a library, the ``v5'' version of the Illinois library (N. J. Bowden et al. 2026, in preparation), of GRMHD simulations of \s{}. In particular, we examine both standard and normal evolution (SANE) magnetically arrested disks (MADs) models for which the original fluid simulations were performed for black holes with spins in the set \{-0.9375, -0.75, -0.5, -0.25, 0, 0.25, 0.5, 0.75, 0.9375\}. We consider images ray traced from these fluid simulations with radiative transfer assuming a small collection of post-processing parameters: viewing inclinations of either $30^\circ$ or $50^\circ$, electron heating parameter $R_{\rm high}$ of $40$ or $160$ \citep{Mosci_2016}, and a mass fixed at $4.1\times10^6$ $M_\odot$. For more details on the general relativistic radiative transfer and ray-tracing pipeline, see \citet{Wong_2022}.

The library was generated in the fast-light approximation, in which all rays in a given observer frame sample a single fluid snapshot, rather than the screen-dependent distribution of emission times used in slow-light transfer. We do not expect this approximation to dominate the unresolved structure functions considered here, because we restrict to moderate inclinations ($i=30^\circ$ and $50^\circ$), for which the integrated fast--slow mismatch is modest \citep{RojasPaternina2026}. In addition, as shown in the previous section, higher-order image contributions affected by the light travel time delay enter as subleading corrections to the direct-image structure function.

We evaluate the structure functions of unresolved light curves from each model. These light curves are sampled at a minimum cadence of $5$ gravitational time units, approximately $100$ seconds apart for \s{}. While the light curves vary in duration depending on the available ray-traced window, the shortest is $15,000$ gravitational time units. As a result, we expect highly converged sampling of short time lags, of particular interest to this work.

We show the structure functions colored by the MAD or SANE state of the disk in the left panel of \autoref{fig:GRMHD}. We first see that SANE disks have a characteristically lower structure function at lower time lags while finding generic agreement with a feature first observed in a single model by \citet{Murchikova_2022}, namely that SANE simulations have a similar structure function log slope as MADs for very short time lags, but quickly bend to a shallower slope. We quantify these differences by computing the structure function power law index $(\alpha)$ using a sliding window of two adjacent points as
\begin{align}
    \alpha_n = \frac{\log( \mathcal{SF}_{n+1}/\mathcal{SF}_{n})}{\log(\Delta t_{n+1}/\Delta t_n)},
\end{align}
where $\mathcal{SF}_{n}=\mathcal{SF}(\Delta t_n).$
In the right panel of \autoref{fig:GRMHD}, we show the power law index $\alpha$ as a function of the initial time lag at which it is evaluated. We find that SANE models bend from a log slope initially just above $0.8$ to a value between $0.5$ and $0.7$ on timescales between 25 and 50 gravitational times, approximately $500$ to $1,000$ seconds for \s{}. We show the computed structure function for the joint \s{} data set in green in both panels, and find two results. First, the magnitude of the \s{} structure function is nearly everywhere lower than both MAD and SANE models, suggesting both model sets are holistically more variable than \s{} in reality. However, the slope of the observed structure function tends to obey the MAD trend.

\section{Discussion}
\label{sec:discussion}

In this work, we have examined the phenomenology of the structure function for (quasi)stationary random processes relevant to black hole accretion flows. We have shown that the time-delayed, demagnified secondary image of the accretion flow can leave an imprint in the structure function that depends sensitively on the spectrum of the underlying turbulence, echoing previous studies of light curve autocorrelations. We have further shown that the tendency of hot accretion flows to ``window'' intrinsic evolutionary timescales to a restricted band, made explicit in our \texttt{inoisy}+\texttt{AART} experiments, can produce kinks or broad knees in the observed structure function. Lastly, using a library of $230$ GHz GRMHD movies of the \s{} accretion flow, we have shown that SANE simulations bend to relatively shallow structure-function slopes on $\sim 500$ s timescales in a way that is not borne out in observations of the real \s{} light curve, further suggesting that \s{} is more consistent with MAD-like dynamics on these lags.

Prior work by \citet{Murchikova_2022} showed hints of the difference between MAD and SANE power-law index evolution at moderate time lags in three models, including one strongly magnetized disk fed by stellar winds \citep{Ressler_2019, Ressler_2020, Ressler_2023}. Our work shows that these signatures generalize to a broader collection of MAD and SANE disks, including a wider variety of emitting geometries, black hole spins, and post-processing prescriptions.

Taken together, these results show that the structure function is a diagnostic that connects three levels of interpretation: the stochastic spectrum of the emitting plasma, the relativistic transfer function that selects the observed mixture of radii and image orders, and the magnetic state of the accretion flow in GRMHD. Because the observational structure function is built from heterogeneous $230$ GHz measurements drawn from epochs between 2009 and 2021, rather than from a single continuous monitoring campaign, this comparison emphasizes lag-dependent slopes and broad curvature rather than dataset-specific amplitudes (which visibly vary by factors of $\sim2$ in Figure \ref{fig:ObsData}). The main implication is that the morphology of the \s{} structure function contains physically useful information even when its absolute normalization is affected by observational noise or imperfect model amplitudes. Conversely, the same hierarchy cautions against interpreting any single kink, knee, or slope as a unique spacetime signature without simultaneously modeling the intrinsic variability prescription.

Throughout this work, we have not treated the absolute magnitude of the structure function as a primary observable for the study of the Galactic Center. Many of the tools we use in this work most naturally align with a scale-free, effectively optically thin representation of the turbulence in the disk. In reality, variability in quantities such as number density can cause the emitting region to move between optically thin and optically thick states over a light curve or between observations. When \s{} is brighter and more optically thick, short time lag variability is likely suppressed. Moreover, the structure function scales with the standard deviation of the observed light curve; as noted extensively in EHT analyses of GRMHD simulations in comparison with \s{} observations, the fractional variability of many GRMHD models greatly exceeds that observed in \s{} \citep{SgrA_PaperV}. Resolving this discrepancy remains a goal for non-ideal GRMHD modeling approaches \citep[see, e.g.,][]{Dhruv_2025}. {More fundamentally, the interpretation of long-lag variability with simulations is limited in part by small initial torus sizes. As GRMHD simulation sizes and durations grow to include many days for Sgr~A* (or many tens of thousands of gravitational times), they can more feasibly treat larger initial tori \citep[see, e.g.,][]{Cho_2024}, allowing safer interpretation of longer monitoring periods in Sgr A*.} As these models advance, structure function analyses will remain critical for comparing the predicted variability power spectrum to the irregularly sampled light curves available in real data. 

\acknowledgments{}

We are grateful to our journal referee for thorough and thoughtful comments on our manuscript. We also thank A. Joshi and C.~Gammie for providing us access to the Illinois library, and to M. Wielgus and G. Bower for providing EHT light curves.

We acknowledge financial support from the National Science Foundation (AST-2307887). This work was supported by the Black Hole Initiative, which is funded by grants from the John Templeton Foundation (Grant 62286) and the Gordon and Betty Moore Foundation (Grant GBMF-8273) - although the opinions expressed in this work are those of the author(s) and do not necessarily reflect the views of these Foundations.

The original GRMHD simulation work was supported in part by NSF grant AST
20-34306. This research is part of the Delta research
computing project, which is supported by the National
Science Foundation (award OCI 2005572), and the State
of Illinois. Delta is a joint effort of the University of Illinois at Urbana-Champaign and its National Center for Supercomputing Applications. This research used resources of the Oak Ridge Leadership Computing Facility at the Oak Ridge National Laboratory, which is supported by the Office of Science of the U.S. Department of
Energy under Contract No. DE-AC05-00OR22725. This research was done using services provided by the OSG Consortium, which is supported by the National Science Foundation awards 2030508 and 2323298.

This work used Delta at NCSA through allocation PHY250091 from the Advanced Cyberinfrastructure Coordination Ecosystem: Services $\text{\&}$ Support (ACCESS) program, which is supported by U.S. National Science Foundation grants $\text{\#2138259}$, $\text{\#2138286}$, $\text{\#2138307}$, $\text{\#2137603}$, and $\text{\#2138296}$.
This work used computing resources provided by Northwestern University and the Center for Interdisciplinary Exploration and Research in Astrophysics (CIERA). This research was supported in part through the computational resources and staff contributions provided for the Quest high performance computing facility at Northwestern University, which is jointly supported by the Office of the Provost, the Office for Research, and Northwestern University Information Technology.
This paper makes use of the following ALMA data: ADS/JAO.ALMA\#2015.1.00311.S, \#2016.1.00870.S, \#2017.1.00995.S, \#2018.1.01124.S, \#2019.1.01559.S. ALMA is a partnership of ESO (representing its member states), NSF (USA) and NINS (Japan), together with NRC (Canada), MOST and ASIAA (Taiwan), and KASI (Republic of Korea), in cooperation with the Republic of Chile. The Joint ALMA Observatory is operated by ESO, AUI/NRAO and NAOJ. The National Radio Astronomy Observatory is a facility of the National Science Foundation operated under cooperative agreement by Associated Universities, Inc.

ChatGPT v5.5 and Microsoft Copilot were used to accelerate the structure function calculation code on long time series. The output was verified to be numerically identical to the original human-written code. The authors take full responsibility for all results presented in this Letter.

\bibliography{references}

\end{document}